\documentclass[intlimits,twoside,a4paper]{article}

\usepackage{amsmath,amssymb}
\usepackage{graphicx}

\usepackage[T2A]{fontenc}
\usepackage[cp1251]{inputenc}

\usepackage[eqsecnum]{cmpj2}

\issue{2014}{17}{4}{43601}
\doinumber{10.5488/CMP.17.43601}

\title[Domain-wall formation in Ar-Kr submonolayer films on graphite]
{The mechanism of domain-wall structure formation in Ar-Kr submonolayer films on graphite}

\author[A. Patrykiejew, S. Soko{\l}owski]{A. Patrykiejew, S. Soko{\l}owski}

\address{Department for the Modelling of Physico-Chemical
Processes, Faculty of Chemistry, MCS University, 20031 Lublin, Poland}

\date{Received July 18, 2014, in final form October 23, 2014}
\authorcopyright{A. Patrykiejew, S. Soko{\l}owski, 2014}

\begin{document}

\newcommand{\mqu}{\mbox{\bf q}}
\newcommand{\mbr}{\mbox{\bf r}}

\maketitle

\begin{abstract}

Using Monte Carlo simulation method in the canonical ensemble, we have studied the
commensurate-incommensurate transition in two-dimensional finite mixed clusters of Ar and Kr
adsorbed on graphite basal plane at low temperatures. It has been demonstrated that the transition
occurs when the argon concentration exceeds the value needed to cover the peripheries of the cluster.
The incommensurate phase exhibits a similar domain-wall structure as observed in pure krypton films
at the densities exceeding the density of a perfect $(\sqrt{3}\times\sqrt{3})R30^\circ$ commensurate
phase, but the size of commensurate domains does not change much with the cluster size. When
the argon concentration increases, the composition of domain walls changes while the
commensurate domains are made of pure krypton.
We have constructed a simple one-dimensional Frenkel-Kontorova-like model that yields the results being
in a good qualitative agreement with the Monte Carlo results obtained for two-dimensional systems.



\end{abstract}


\section{Introduction}
\label{INTRODUCTION}

The commensurate-incommensurate (C-IC) transitions in adsorbed films have been experimentally observed
in a variety of
systems \cite{cicex1,cicex2,cicex3,cicex4,cicex5,cicex6,cicex7,cicex8,cicex9,cicex10,cicex11,cicex12,cicex13,cicex14,cicex15}
and studied by theoretical methods \cite{cict1,cict2,cict3,cict4,cict5,cict6,cict7,cict8,cict9,cict10,cict11,cict12}
and computer simulations \cite{cicsym1,cicsym2,cicsym3,cicsym4,cicsym5,cicsym6,cicsym7,cicsym8,cicsym9,cicsym10}.

In a large number of systems, the incommensurate phase forms a collection of finite commensurate regions (domains)
separated by domain walls \cite{cict6}. The domain wall networks appear when the C phase is degenerated,
i.e., when the atoms (or molecules) in the C phase occupy one of several equivalent lattices. The stability of IC
phases with domain walls networks is considerably enhanced by large entropic effects \cite{cict1,cict6}.
The most prominent example of such a system is the krypton monolayer adsorbed on the graphite
basal plane \cite{krgr1,krgr2,krgr3,krgr4}.
At low temperatures, the adsorption of krypton leads to the formation of the three-fold degenerate
commensurate $(\sqrt{3}\times\sqrt{3})R30^\circ$ $(\sqrt{3}\times\sqrt{3})R30^\circ$ structure,
in which every krypton atom occupies one of every three hexagons of the graphite surface, has the lattice
constant $\sqrt{3}$ times the graphite lattice constant ($a_\textrm{gr}=2.46$~{\AA}) and is rotated with
respect to the graphite lattice by 30 degrees. Upon the increase of pressure, the commensurate
phase undergoes a transition into the incommensurate phase of higher density, in which
the commensurate domains are separated by walls, and the atoms in neighboring domains
are shifted to a different sublattice \cite{krgr3,gr18,krgr5,gr17}. In this case, the walls are called
heavy and carry all the excess of density. At the densities just above the C-IC transition the domains are very
large and their size gradually decreases when the density increases towards the monolayer completion. This has
been clearly demonstrated by a large scale computer simulation study \cite{cicsym3}.

The C-IC transition has also been experimentally observed in the mixed Kr-Xe and Ar-Xe films on graphite \cite{krxe1,arxe2,arxekrxe,arxe3}. On the other hand, the structure of mixed Ar-Kr films on graphite has not
been experimentally studied yet. The only available experimental results were published by Singleton and Halsey nearly 60 years ago \cite{SinHal} and demonstrated that mixed liquid-like films exhibit complete mixing, independently of
the composition. The only information about the structure and properties of solid-like mixed Ar-Kr films on graphite
have been obtained using computer simulation methods \cite{langmuir2,Ar-Kr-my}. In particular, it
has been demonstrated that already
at submonolayer coverages and at sufficiently low temperatures, the addition of argon into krypton
film triggers the C-IC transition, and
leads to the formation of finite clusters in which small commensurate domains consisting of pure
krypton are separated by heavy walls built of argon and krypton \cite{Ar-Kr-my}. The C-IC transition in
one-component finite clusters of Lennard-Jones atoms adsorbed on graphite has also been studied by
Houlrik et al. \cite{cicsym10}. They have used Monte Carlo
method to study the effects of the corrugation potential on the structure of finite systems and have
shown that for a given amplitude of the corrugation potential the presence of free
surfaces enhanced the stability of the C structure.

In this paper we study the mechanism of the C-IC transition of submonolayer mixed films made of argon and krypton
using Monte Carlo simulation methods and appropriately modified one-dimensional (1D) Frenkel-Kontorova (FK) model \cite{FK1,FK-my}. The original FK model handles an infinite chain
of atoms interacting via harmonic potential and subjected to a periodic (sinusoidal) external field at zero temperature.
The FK model can be used to describe the basis features of C-IC transitions \cite{cict2}.
At this point, we should mention that the FK model has been extended to
two-dimensional systems \cite{FK2D-1,FK2D-2} to mixtures \cite{FK-mix1}, systems with
disorder \cite{FK-dis1,FK-dis2} and has also been used to study finite
chains \cite{FK-fin1,FK-fin2,FK-fin3,FK-fin4,FK-fin5}.

In one of our recent works \cite{FK-my}, we  studied the impurity driven commensurate-incommensurate
transitions in one-dimensional
finite systems using the FK model.
We assumed that the  pure chain is commensurate and discussed the effects of impurities located
either only at one end or at two ends of the chain.
It is shown that in both situations the C-IC transition occurs when the
amplitude of the external field experienced by the impurity atoms falls into the region between the
lower and upper threshold values. These limiting values of the amplitude depend upon
the parameters characterizing the interaction between the atoms in the main chain,
the amplitude of external field acting on the main chain atoms
and the interaction between the main chain and the impurities. The number of solitons
(domain walls) in the IC structure were found to depend upon the chain length, and the parameters
describing the interactions in the system. Our findings have been confirmed by Monte Carlo simulation
in two-dimensional systems.

Here, we also considered finite chains of Kr atoms with impurities (Ar atoms) located at the chain
ends as well as in the chain interior.

The paper is organized as follows. In the next section we  presented the two-dimensional model and described the
Monte Carlo method used. The results of Monte Carlo simulation are presented in section~3. Then, in section~4 we
presented the modified 1D FK model and demonstrated that it yields the results qualitatively very similar to
those emerging from the two-dimensional Monte Carlo simulation.

\section{The two-dimensional model and simulation method}

We have considered strictly two-dimensional mixed submonolayer films consisting of Ar and Kr atoms. The
interaction between the atoms is modelled via the truncated (12,\,6) Lennard-Jones potential
\begin{equation}
u_{ij}(r) = \left\{ \begin{array}{ll}
4\varepsilon_{ij}\left[\left(\sigma_{ij}/r\right)^{12} - \left(\sigma_{ij}/r\right)^{6}\right],   & r \leqslant r_\textrm{max} \, ,\\
0,                                                & r > r_\textrm{max}\, ,
\end{array}
\right.
\label{eq:01a}
\end{equation}
where $r$ is the distance between a pair of atoms, the subscripts $i$ and $j$ mark the species Ar and Kr, and
the potential is cut at the distance $r_\textrm{max}=3.0\sigma_{ij}$.
The potential parameters for a pair of unlike atoms are given by the Lorentz-Berthelot relations:
\begin{equation}
\sigma_{ArKr}=\frac{1}{2}\left(\sigma_\textrm{Ar}+\sigma_\textrm{Kr}\right) \qquad  \text{and} \qquad \varepsilon_\textrm{ArKr} = \sqrt{\varepsilon_\textrm{Ar}\varepsilon_\textrm{Kr}}\,.
\label{eq:01b}
\end{equation}

The external field due to the graphite substrate is assumed to be given by \cite{cicsym6,Ar-Kr-my}
\begin{equation}
v(x,y)= -V_{\textrm{b},i}\left\{\cos(\mqu_1\mbr)+\cos(\mqu_2\mbr)+\cos[(\mqu_1-\mqu_2)\mbr]\right\},
\label{eq:06}
\end{equation}
where $\mqu_1$ and $\mqu_2$ are the reciprocal lattice vectors of the graphite basal plane and the
amplitudes $V_{\textrm{b},i}$ ($i=\textrm{Ar}, \textrm{Kr}$) determine the potential barriers between adjacent minima. We
assumed here that $V_{\textrm{b},\textrm{Ar}}^{\ast}= 8.4$~K and $V_{\textrm{b},\textrm{Kr}}^{\ast}= 14.4$~K \cite{Ar-Kr-my}.

The values of parameters entering the potential (\ref{eq:01a})
are given in table~\ref{tab1}.
The graphite lattice constant $a_\textrm{gr}=2.46$~{\AA} is assumed to be a unit of length and $\varepsilon_\textrm{Ar}$
is taken as a unit of energy.
\begin{table}
\begin{center}
\caption{Lennard-Jones parameters for Ar and Kr used in this work\label{tab1}}
\vspace{2ex}
\begin{tabular}{|c|c|c|}
\hline\hline
 $i,j$   &  $\sigma_{i,j}$, {\AA}  & $\varepsilon_{i,j}$, K \\
\hline
Ar,Ar  &   3.4 & 120.0  \\
Kr,Kr  &   3.6 & 171.0  \\
Ar,Kr  &   3.5 & 142.83 \\ \hline\hline
\end{tabular}
\end{center}
\end{table}

The Monte Carlo simulation is carried out in the canonical ensemble \cite{BL2000} for systems of
different (submonolayer) density, of different total number of atoms, $N$, and of different mole fraction
of argon, $x_\textrm{Ar}$, between 0 and 0.55.  We used a standard Metropolis sampling and two types of moves are taken into account, i.e., the translation of a randomly chosen atom by  a randomly chosen vector within the
circle of radius $d_\textrm{max}$ and the identity exchange. The magnitude of $d_\textrm{max}$ is
updated every 1000 Monte Carlo steps (each Monte Carlo step consists of $N$ attempts to move a single atom as well as $N$ attempts to change the identity of randomly chosen atoms) in order to keep the acceptance rate
equal to about 0.3. The equilibration and production runs consist of $5\cdot10^6$ and   $5\cdot10^7$ Monte Carlo
steps, respectively.  Standard periodic boundary conditions were used therein.

\section{The results of Monte Carlo study of Ar-Kr submonolayer films}

Since our study was performed for finite clusters at low temperatures it seems reasonable
to begin with the estimation of the effects of free boundaries on the ground state properties
of finite commensurate patches made of pure krypton.

We  performed Monte Carlo calculations at low temperatures using different starting configurations.
At first, the rectangular simulation cells of the
size $L_x\times L_y=L\times L\sqrt{3}/2$ with $L=30$, 60 and 120, were used. The number of Kr atoms in a
perfect and fully covered $(\sqrt{3}\times\sqrt{3})R30^\circ$ commensurate structure was equal to $N_\textrm{Kr} = L^2/3$.
Then, we modified the
simulation cell by making $L_x$ larger ($L_x=L+\Delta$),
while keeping the number of Kr atoms unchanged, i.e., equal to $N_\textrm{Kr} = L^2/3$. In this way, two
free boundaries running along the $y$-axis were created. Of course, the distance between these two boundaries,
given by $\Delta$, should be larger than the cutoff distance of the
Lennard-Jones potential [equation (\ref{eq:01a})]. The second series of calculations were done for finite hexagonal
clusters of the commensurate phase built of a different number of krypton atoms.
The simulation was carried out at reduced temperatures between $0.01$ and $0.1$ and the
energies obtained were extrapolated to $T=0$. In this way, we estimated the ground state energies.

\begin{figure}[!t]
\centerline{
\includegraphics[scale=0.42]{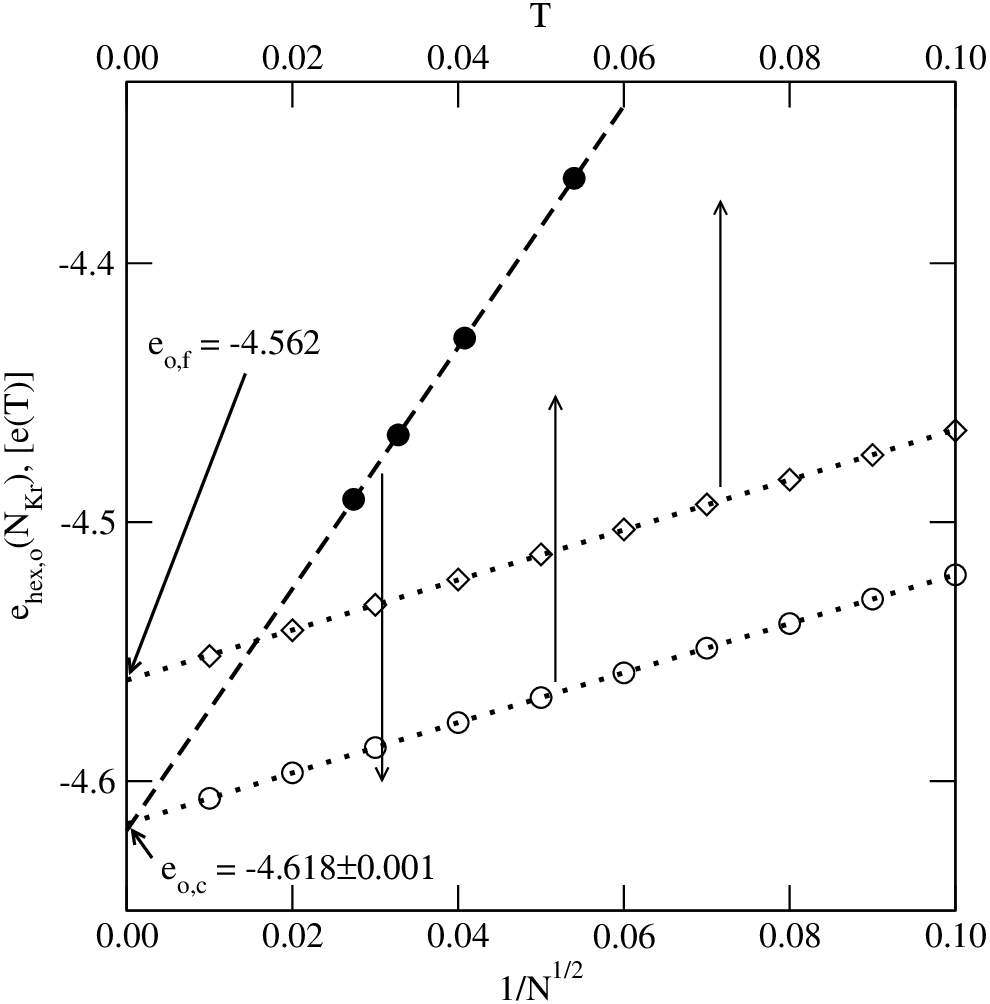}
}
\caption{The temperature changes of the energy (per particle) of fully covered
(open circles) and with free boundaries (open diamonds) commensurate phase obtained using a rectangular
slabs containing 600, 1200 and 1800 atoms and of different total length of the free interface, $L_\textrm{int}=L_y\sqrt{3}$,
with $L_y = 30$, 60 and 90.  The filled circles are the ground state energies of
finite hexagonal patches plotted against $N^{-1/2}$.
\label{fig1}}
\end{figure}

In the case of a fully occupied rectangular box, we did not observe any finite size effects, and
the ground state energy (per particle) of the commensurate phase was estimated to be
equal to $e_{0,\textrm{c}}=-4.617$ (see figure~\ref{fig1}).
Similar calculations carried out for the systems with free boundaries along the $y$-axis led to the ground state
energy equal to $e_{0,\textrm{f}}=-4.562$ (see figure~\ref{fig1}) for three different systems of the size $(60\times 15\sqrt3 )$ ($N_\textrm{Kr}=600$),
$(60\times 45\sqrt 3)$ ($N_\textrm{Kr}=1800$) and $(90\times 30\sqrt3)$ ($N_\textrm{Kr}=1200$).

In the ground state, the excess interfacial energy per unit length due to the presence of free boundaries of the total
length $L_\textrm{int}$ can be calculated as follows:
\begin{equation}
E_\textrm{int}= (E_{0,\textrm{f}}-E_{0,\textrm{c}})/L_\textrm{int}\, ,
\label{eq:3.1}
\end{equation}
where $E_{0,\textrm{f}}=N_\textrm{Kr}e_{0,\textrm{f}}$ and $E_{0,\textrm{c}}=N_\textrm{Kr}e_{0,\textrm{c}}$ are the total energies of the systems
with and without  free boundaries. For the three systems mentioned above, we  obtained the same
value of the excess interfacial energy at $T=0$ equal to about 0.635. This indicates that finite size
effects are negligibly small at very low temperatures.

On the other hand,
the ground state energies of finite hexagonal clusters are bound to show large finite size effects. In such
clusters, the fraction of atoms at the patch boundaries is proportional to $1/\sqrt{N}$.
In order to extrapolate the results to the limit $N_\textrm{Kr}\rightarrow\infty$, we made the plot
(given in figure~\ref{fig1}) of the ground state energies (per atom) for
the clusters of different size $e_\textrm{hex,0}(N_\textrm{Kr})$ against  $1/\sqrt{N}$ and extrapolated the results to
$1/\sqrt{N}=0$. In this limit, the boundary effects should vanish and the energy for $N_\textrm{Kr}\rightarrow\infty$
should be equal to $e_{0,\textrm{c}}$.
The obtained value is equal to $-4.169$ and agrees very well with the value of $e_{0,\textrm{c}}$ obtained
for the fully covered surface. The above presented results demonstrate that our simulation method gives
reliable results.

It was demonstrated in our earlier work \cite{Ar-Kr-my}, that finite submonolayer clusters
made of the mixture of Ar and Kr atoms exhibit domain-wall structures, in which the commensurate
domains made of krypton are separated by the domain walls of varying composition. Here,
we considered such systems in a more systematic way, aiming at the determination of conditions
necessary for the domain-wall structure to appear.

\begin{figure}[!t]
\centerline{
\includegraphics[scale=0.4]{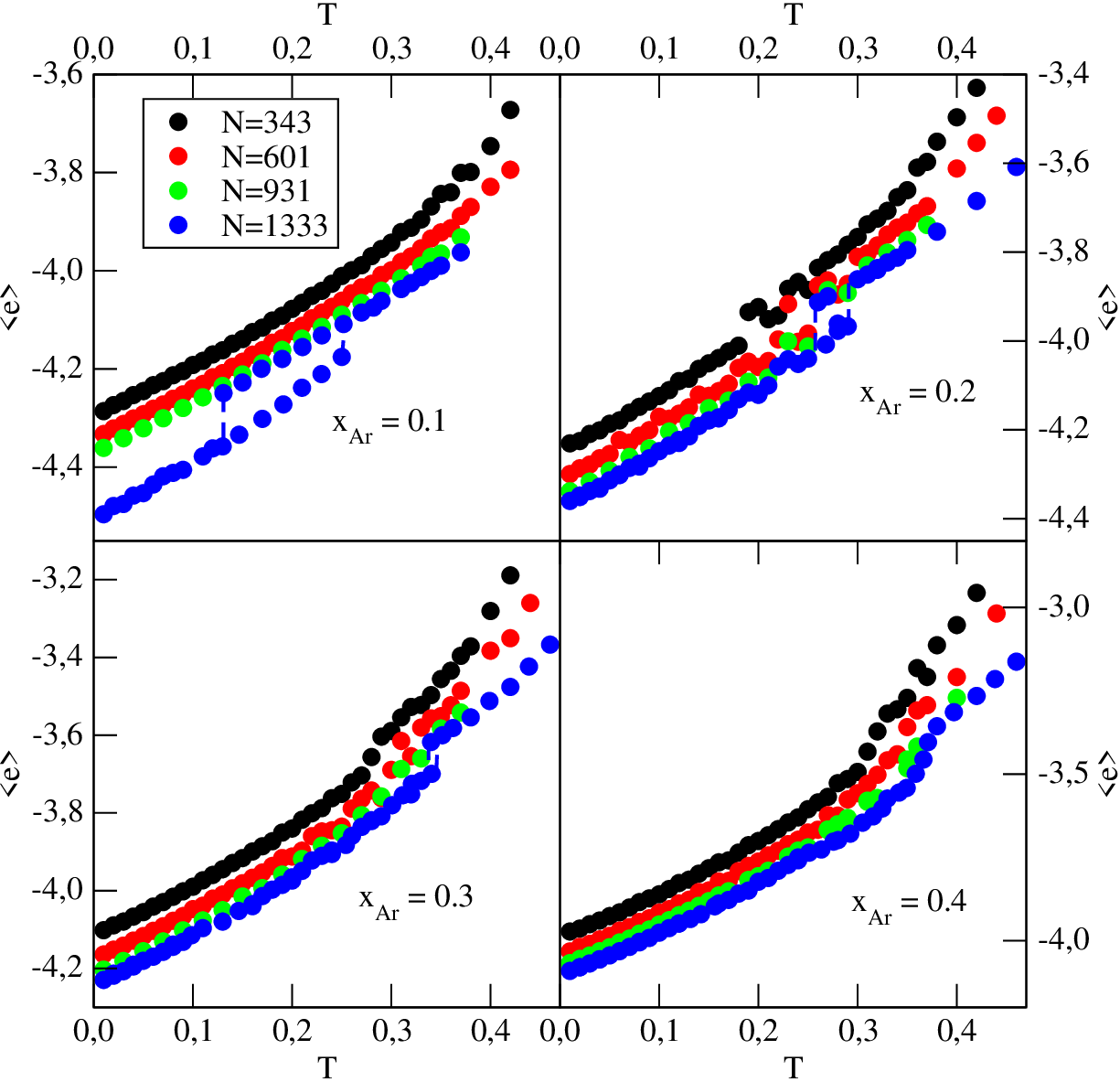}
}
\caption{(Color online) The temperature changes of the potential energy (per particle) obtained for finite
hexagonal patches of different size and of different argon mole fraction. The vertical blue lines mark
the locations of transition between the mixed commensurate structure (stable at high temperatures) and the domain-wall
structure (stable at low temperatures) in the largest clusters of $N=1333$ atoms, obtained from annealing and freezing runs.\label{fig2}}
\end{figure}

\begin{figure}[!t]
\centerline{
\includegraphics[scale=0.35]{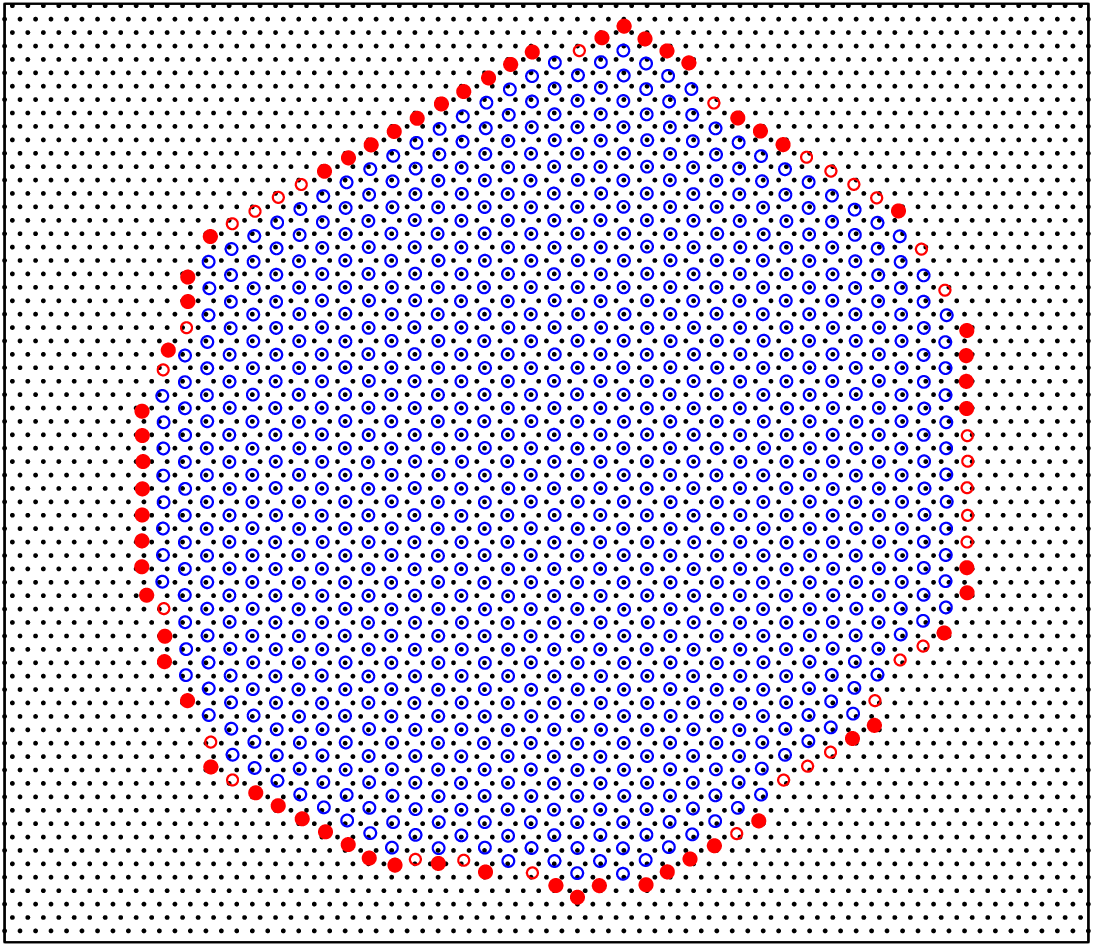}
\includegraphics[scale=0.35]{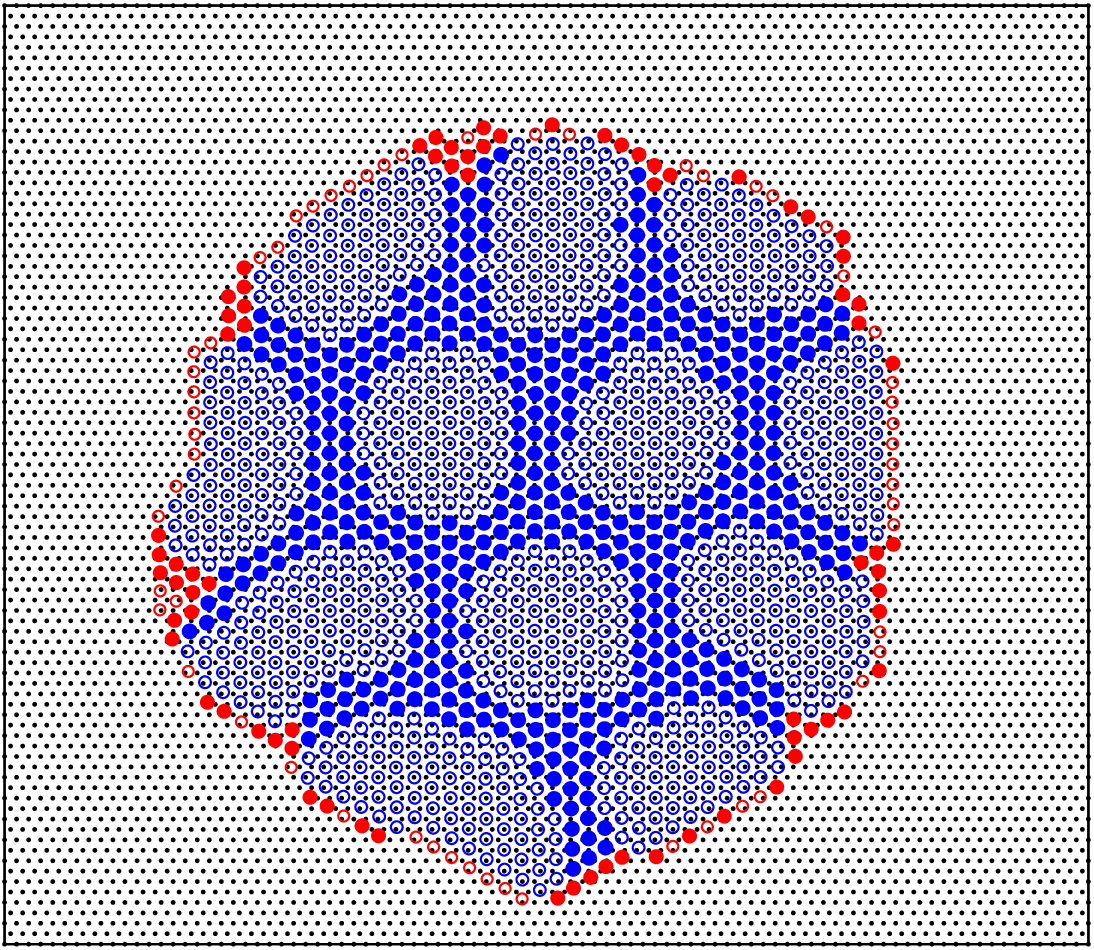}
}
\caption{(Color online) The examples of configurations for the finite patches with $x_\textrm{Ar}=0.1$ at $T=0.02$ consisting of 931
(part a) and 1333 (part b) atoms. Blue and red circles mark krypton and argon, respectively. The open circles correspond
to the commensurate atoms with the order parameter
$\phi(\mbr_i)\geqslant 0$, while the filled circles mark the incommensurate atoms with $\phi(\mbr_i)<0$.\label{fig3}}
\end{figure}

The Monte Carlo simulations were carried out for a series of hexagonal clusters
of a different total number of atoms equal to 343, 601, 931 and 1333
and of also different argon mole fraction between 0.08 and 0.55. The calculations were performed over a rather
wide range of reduced temperatures between 0.01 and 0.45.
Figure~\ref{fig2} shows the temperature changes of the potential energy (per particle) in the systems of different
cluster size and of different composition. Only in the case of $x_\textrm{Ar}=0.1$ and the clusters with $N$ up
to 931, the energy smoothly changes with temperature, and the inspection of snapshots showed that
all krypton atoms form a commensurate patch while all argon atoms are located along the patch boundary
[see figure~\ref{fig3}~(a)].

In order to distinguish the commensurate and incommensurate atoms, we used the following
order parameter \cite{cicsym4,cicsym10,Ar-Kr-my}
\begin{equation}
\phi(\mbr)=\cos(\mqu_1\mbr)+\cos(\mqu_2\mbr)+\cos[(\mqu_1-\mqu_2)\mbr]\,,
\label{eq:local}
\end{equation}
and assumed that the atom is commensurate (incommensurate) when $\phi>0$ ($\phi\leqslant 0$).

In all other
cases, the domain-wall structures were observed at
sufficiently low temperatures.
Figure~\ref{fig3}~(b) shows the example of a snapshot recorded at $T=0.02$ and $x_\textrm{Ar}=0.1$ for the larger patch made of 1333 atoms.
It can be clearly seen that unlike in the smaller cluster of $N=931$ atoms, the cluster with $N=1333$
exhibits the domain-wall structure, and the
walls are made nearly entirely of krypton atoms. The argon contributes to the walls only in the areas close to the patch boundaries. The patch boundaries are covered by a single layer of argon atoms. The structure of the IC
phase is qualitatively similar to the IC phase in pure
krypton monolayers of the density exceeding that of a fully filled C structure.
However, the mechanism leading to the commensurate-incommensurate transition is quite different in
both situations. In the case of a fully filled monolayer, the density excess over the value corresponding
to a perfect C phase triggers the formation of domain walls.
Upon the increase of density, the size of commensurate domains gradually decreases \cite{cicsym3,cicsym11}.
Ultimately, a dense IC phase without commensurate domains is formed close to the monolayer completion.

\begin{figure}[!b]
\centerline{
\includegraphics[scale=0.35]{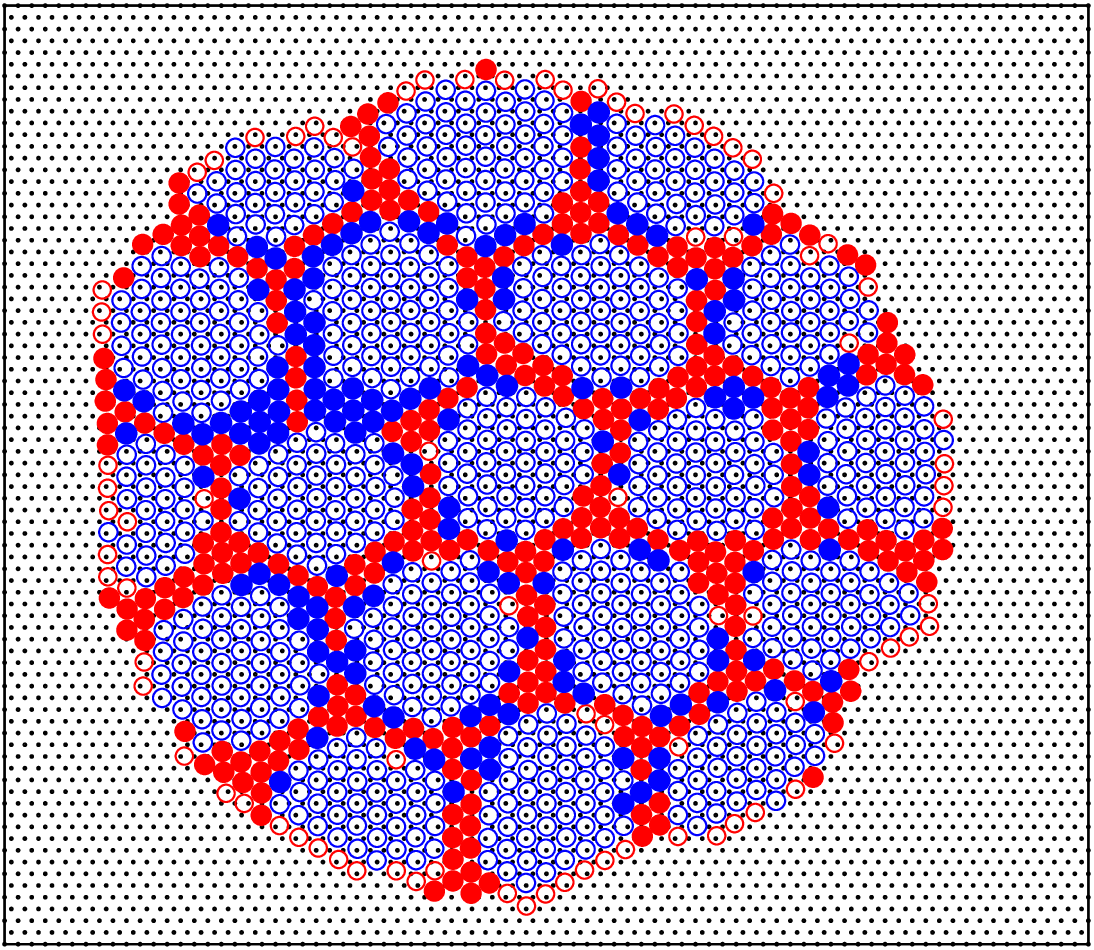}
}
\caption{(Color online) The example of configuration for the cluster with $N=1333$ atoms and $x_\textrm{Ar}=0.3$
at $T=0.06$. The labeling of atoms is the same as in figure~\ref{fig3}.\label{fig4}}
\end{figure}

\begin{figure}[!t]
\centerline{
\includegraphics[scale=0.38]{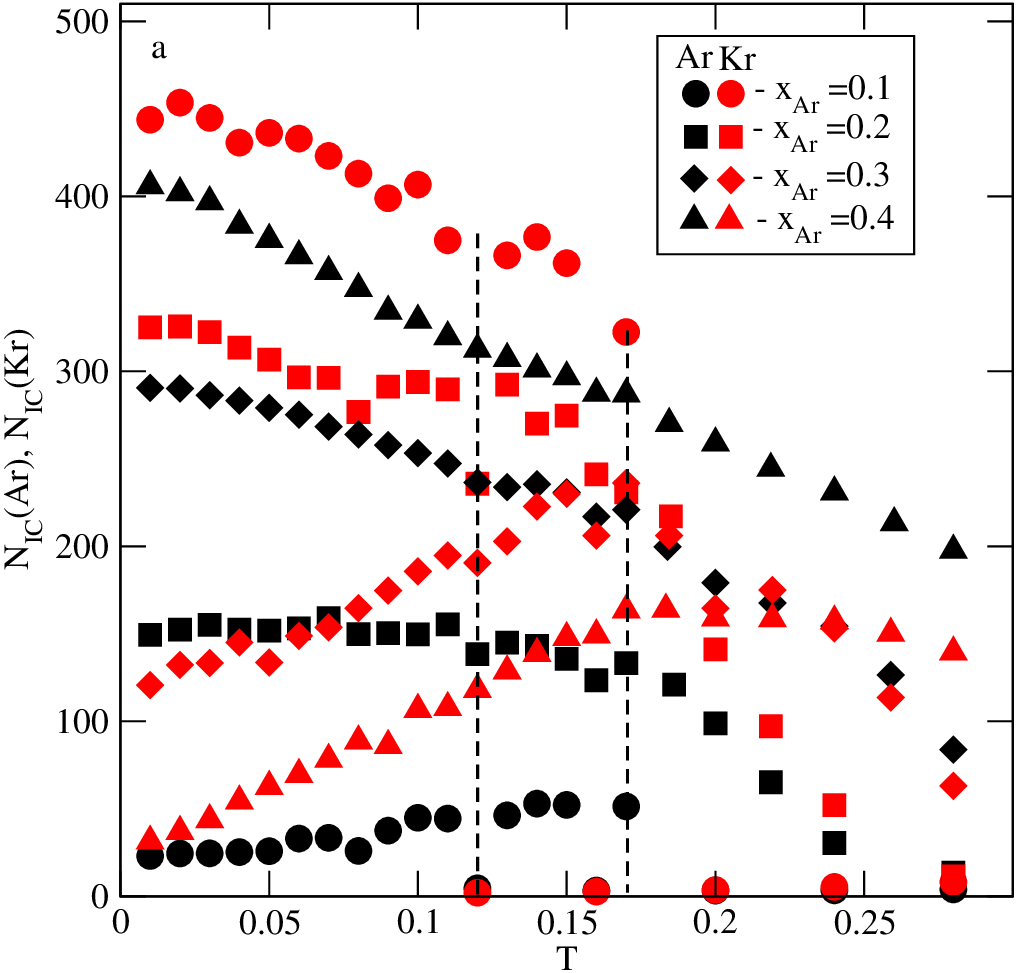}
\hspace{5mm}
\includegraphics[scale=0.35]{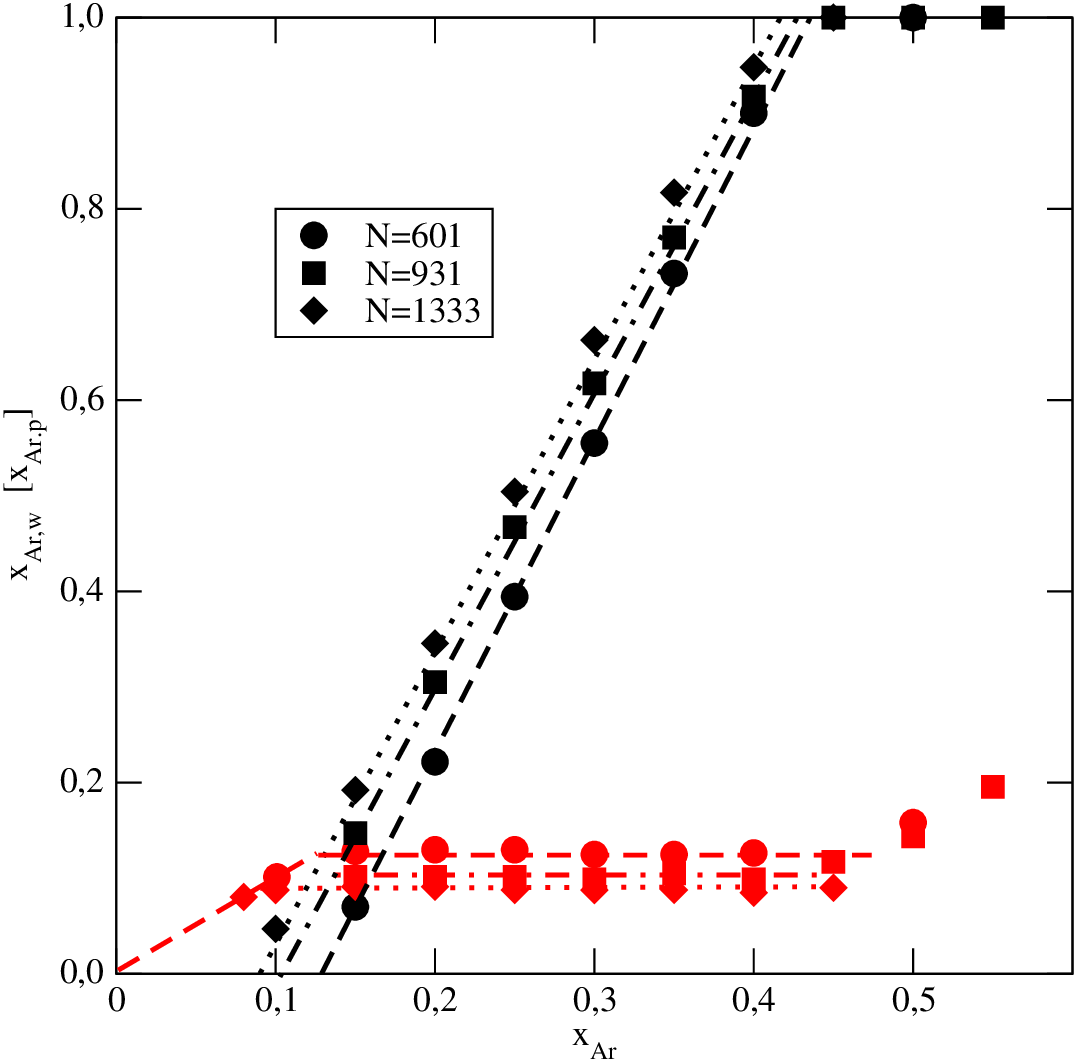}
}
\caption{(Color online) (part a) The temperature changes of the numbers of argon and krypton atoms within the walls in finite
clusters of $N=1333$ for the
systems with different argon mole fraction equal 0.1 (circles), 0.2 (squares) 0.3 (diamonds) and 0.4 (triangles).
Black and red symbols mark the numbers of argon and krypton atoms, respectively. (Part b). The mole fraction of
argon within the wall (black symbols) and the mole fraction of argon atoms at the
patch boundaries (red symbols) against the total mole fraction of argon atoms obtained for clusters of
different total number of atoms (given in the figure). The lines serve only as a guidance.}
\label{fig5}
\end{figure}

In finite mixed clusters the atoms close to the boundaries are
strained and the magnitude of this strain depends on the  number of argon atoms in the system.
It is shown that the excess interfacial energy in pure krypton clusters is positive
and it induces a force normal to the interface and pointing towards the patch interior. Evidently, this
force is too weak to lead to the  C-IC transition in the pure krypton clusters. The
presence of a layer of argon atoms at the boundaries causes the interfacial energy to be considerably
higher and hence induces a considerably larger force, which is sufficient to compress the patch
and to trigger the C-IC transition. The number of argon atoms
necessary to cover the entire patch boundary is proportional to the total length of the interface, which
in turn is proportional to $N_\textrm{Kr}^{1/2}$. For a small and fixed argon mole fraction in the system,
the concentration of argon along the interface increases with the cluster size. Figure~\ref{fig3}~(a) shows that
in the case of $N=931$ and $x_\textrm{Ar}=0.1$, argon atoms hardly cover the entire interface. On the other
hand, when $N=1333$ [figure~\ref{fig3}~(b)], the entire interface is densely covered by argon and there is
some excess of argon atoms which try to penetrate the patch and trigger the development of domain walls.
When the argon concentration increases, the walls become mixed and for a large argon mole fraction
are predominantly built of argon (see figure~\ref{fig4}).
We  studied the changes of the wall composition resulting from the changes of argon concentration
in the system for the clusters of different size with $N=601$, 931 and 1333.
We did not take into account the incommensurate atoms with less than 5 nearest neighbors. Such atoms
are located along the patch boundaries, and the inspection of snapshots (see figure~\ref{fig3} and \ref{fig4}) allowed
us to assume that there is a single atomic layer of incommensurate atoms at the patch boundaries.
Using such a procedure we were able to monitor the changes of the number of argon and krypton
atoms incorporated within the walls and to estimate the argon concentration within the wall and
along the patch boundaries. We performed a series of Monte Carlo simulations over a certain
temperature range between 0.01 and 0.3 for different cluster sizes and different argon mole
fraction in the film. The examples of temperature changes of the wall composition are given
in figure~\ref{fig5}~(a), which shows the numbers of argon and krypton atoms forming the walls in the clusters
of 1333 atoms and different argon mole fraction. The results demonstrate that for $x_\textrm{Ar}=0.1$,
the patch is commensurate at sufficiently high temperatures, and undergoes a transition
to the incommensurate structure at $T=0.14\pm 0.02$. The transition seems to be of the first order as
suggested by rather large hysteresis along the freezing and annealing runs. In the case of $x_\textrm{Ar}=0.2$, the
transition occurs at higher temperature of about  $T=0.20\pm 0.02$ and is rounded. For the patches with
a higher argon concentration, the incommensurate structure occurs over the entire range of temperatures studied.
It is also evident that upon the increase of argon concentration in the film, the composition
of walls changes from krypton rich to argon rich.
Having the temperature changes of the wall composition we could estimate the ground state
behavior of the systems studied.


The results are given in figure~\ref{fig5}~(b), which shows the changes of the argon mole fraction within the walls and
at the patch boundaries versus the total argon mole fraction in the system.
These results demonstrate that when the cluster size increases, the onset of the C-IC transition
occurs at a lower argon concentration in the film. In the cluster with $N=601$, it occurs when $x_\textrm{Ar}\approx 0.125$,
in a larger cluster consisting of 931 atoms it starts when $x_\textrm{Ar}$ is slightly higher than 0.1, while in the largest
cluster of $N=1333$ the transition is found already when $x_\textrm{Ar}\approx 0.09$. Of course, when the
argon mole fraction is lower than the limiting value necessary to trigger the C-IC transition, all argon atoms
are located along the patch boundaries.
The argon mole
fraction within the walls increases linearly with the argon mole fraction in the film and for sufficiently large
argon concentration the walls  are
entirely formed by argon. A further increase of the argon
mole fraction leads to a gradual decrease of the size of krypton commensurate
domains \cite{Ar-Kr-my}.

The inspection of configurations showed that when the walls are
predominantly made of krypton atoms (at low argon concentrations) they are wider than the walls
formed in the films with high argon concentration. The domain walls made of  argon atoms only were
observed to usually
consist of just two rows of argon atoms and can be classified as heavy walls. The neighboring domains are shifted by
the displacement vector of the length
equal to $a_\textrm{gr}$ \cite{cict6}. The results obtained for different cluster sizes showed that the
domains are rather small, and usually contain up to 10 commensurate krypton atoms along the line joining  opposite walls.
This is quite different from the already mentioned pure krypton IC phase, in which  the domains are very large
close to the C-IC transition and their size gradually decreases when
the density becomes higher.

We  also studied semi-infinite systems consisting of the rectangular commensurate domain of the size
$L_x\times L_y$ with periodic
boundary conditions applied along the $x$ and $y$-axes, but with free interfaces running along the $y$-axis.
The simulation cell was of the size $L_x=60+\Delta$, with $\Delta=30$, and $L_y=30\sqrt{3}/2$.

Starting from a perfect C structure, we  estimated the ground state energies for
three different configurations with the fixed numbers of Ar and Kr atoms equal to $N_\textrm{Ar}= 60$
and $N_\textrm{Kr}=540$. The simulations were carried out at low temperatures, between 0.01 and 0.1,
using two types of algorithms, with and without the identity exchange attempts.

Three different systems were considered. In the first (i), two rows of krypton atoms adjacent to each
boundary running along the $y$-axis were replaced by argon atoms. The second system (ii)
contained only single rows
of argon atoms at both free boundaries and  the rest of Ar atoms formed a single
wall consisting of two rows of atoms inside the C patch at a certain distance $r_w$ from one of
the interfaces.
The third system (iii) also consisted of single rows of argon atoms along the two free boundaries,
but the rest of Ar atoms formed two walls, each made of one row of atoms located
at the distance $r_w$ from the closer free interface. The calculations showed that in the systems
(ii) and (iii) energy becomes independent of $r_w$ as soon as $r_w$ is larger
than the assumed cutoff distance $r_\textrm{max}$. The results given below were obtained for $r_w=10$.

The simulation without identity exchange attempts demonstrated that the system (ii) has a lower
ground state energy (per particle) ($e_\textrm{i}\approx -4.450$) than the systems (i) ($e_\textrm{ii}\approx -4.425$)
and (iii) ($e_\textrm{iii}\approx -4.398$).
Moreover, the entire krypton patch in the system (i) was observed to form the C structure at low
temperatures, while the vast majority of argon atoms were displaced to incommensurate positions
due to a certain contraction resulting from the presence of free interfaces.
In the case of system (ii), argon atoms were observed to form an incommensurate wall with
the displacement vector between the commensurate domains at both sides of the wall corresponding
to a heavy wall.  The structure of system (iii) was found to be commensurate, just the same
as in the case of system (i)
Another Monte Carlo run, in which the identity exchange attempted was taken into account,
performed for the system (i) demonstrated that already at very low temperatures the system
structure spontaneously changes to that corresponding to the system (ii). Moreover, the two single rows
of argon atoms inside the patch of the system (iii) were observed to merge into a single
heavy wall consisting of two rows of argon atoms.
We also performed a run using the algorithm involving the identity
exchange attempts and the staring configuration with three rows
of argon atoms at each free boundary. In this case, we observed the formation of two heavy walls
running along the y-axis and made of two rows of argon atoms again.

We  calculated the ground state energies  of perfectly commensurate ($E_{\textrm{c},k}$) and relaxed ($E_{\textrm{rel},k}$)
structures for the
systems $k=$(i), (ii) and (iii) as well as the energy gain (per the unit length of  free interfaces) due to the relaxation
\begin{equation}
e_{\textrm{int},k}= (E_{\textrm{rel},k}-E_{\textrm{C},k})/L_\textrm{int}\,.
\end{equation}

The energies (per atom) of perfectly ordered commensurate structures  are very similar and equal to $e_{\textrm{C},1}=-4.3748$,
$e_{\textrm{C},2}= -4.3699$ and $e_{\textrm{C},3}=-4.36707$. On the other hand, the energy gains due to relaxation are considerably
different and equal to $e_{\textrm{int},1}\approx -1.16$, $e_{\textrm{int},2}\approx -1.85$ and $e_{\textrm{int},3}\approx -0.76$.

\begin{figure}[!h]
\centerline{
\includegraphics[scale=0.44]{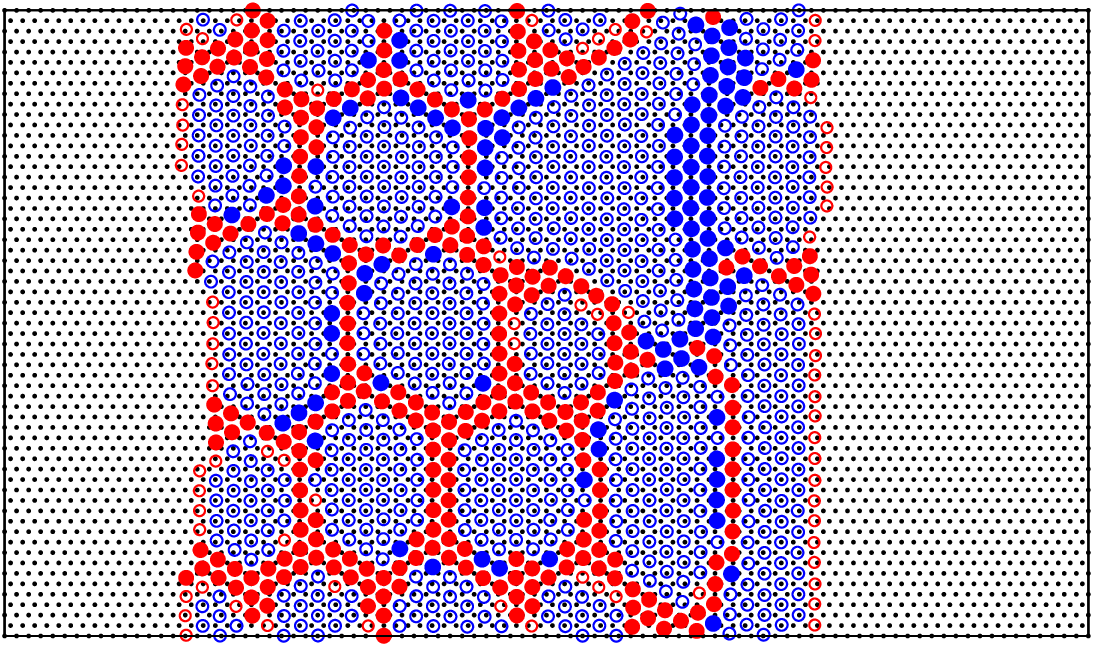}
}
\caption{(Color online) The example of configuration for the rectangular slab of the size $60\times 30\sqrt{3}$ with $x_\textrm{Ar}= 0.2$
at $T=0.03$.
Blue and red circles mark krypton and argon respectively. Circles with thin lines correspond
to commensurate atoms with the order parameter
$\phi(\mbr_i)\geqslant 0$, while circles with thick lines are for incommensurate atoms with $\phi(\mbr_i)<0$.\label{fig6}}
\end{figure}

The above presented results demonstrate that the formation of heavy walls consisting of two rows of argon
atoms inside the krypton patch leads to the most stable structure.
It should be also noted that simulation for the systems with larger argon concentrations  lead to
the development of networks of heavy walls like that given in figure~\ref{fig6}, rather than to a series of parallel walls
running along the $y$-axis. This demonstrates that wall crossings contribute to the system stability and suggests
that the wall crossing energy is negative. Of course, at finite temperatures the network of hexagonal walls
is also stabilized by entropic effects, in particular, by the so-called breathing entropy \cite{cict7,cict8,vill}.

Concluding this section, we should mention that from our earlier study \cite{Ar-Kr-my} it follows that upon the
increase of temperature the domain walls structure disappears and the mixed commensurate phase is
formed at the temperatures below the melting point.

\clearpage

\section{A simple one-dimensional Frenkel-Kontorova model}

In this section, we propose a simple modified one-dimensional Frenkel-Kontorova model \cite{FK1,FK-my}
that leads to qualitatively similar results as those discussed in the previous section.

Taking into account that the domain-wall structures that appear for submonolayer
coverages are stable only at very low temperatures, we consider finite chains at zero temperature.

In general, the energy of the chain consisting of $N$ atoms of two different species A and B can be written as
\begin{equation}
E = \frac{1}{2}\left\{\sum_{i=1}^{N-1}K_{i,i+1}\left[x_{i+1}-x_{i}-b_{i,i+1}\right]^2 + \sum_{i=1}^Nv_{i}\left[1-\cos(2\pi x_i/a)\right]\right\}.
\label{eq:Kon1}
\end{equation}
In the above, we  assumed that the interaction between a pair of nearest neighbors is harmonic and
characterized by the force constant $K_{i,i+1}= K(\textrm{A,A})$, $K(\textrm{A,B})$ or $K(\textrm{B,B})$ and the equilibrium
distance $b_{i,i+1} = b(\textrm{A,A})$, $b(\textrm{A,B})$
or $b(\textrm{B,B})$, depending on the composition of the pair. The second sum in equation (\ref{eq:Kon1})
represents the contribution to the potential energy due to the periodic substrate field, with the
distance between adjacent minima equal to $a$, and the amplitude of the external field $v_i$ is equal to
$v_\textrm{A}$ or $v_\textrm{B}$.

In order to find the equilibrium configuration of such a mixed chain of the length $N$, one needs to specify
the numbers of atoms A and B and their positions along the chain, and then to minimize the energy
with respect to the set of $\{x_i\}$.

Here, we considered a few situations that are supposed to
mimic the behavior of the previously discussed two-dimensional systems.
We  assumed that the component A is Kr-like, while the component B is Ar-like.
The equilibrium configuration of the chain made of atoms A
only should correspond to the commensurate structure. On the other hand, the chain made of only B
atoms should lead to an incommensurate floating structure. Assuming that the elastic constant
$K(\textrm{A,A})$ is the unit of energy and the substrate lattice constant $a$ is the unit of length,
we found out that the model with the parameters $v_\textrm{A}=0.0055$ and $b(\textrm{A,A})=1.9$ describes reasonably well the commensurate structure. Note that by taking $b(\textrm{A,A})=1.9$ we assumed that in the
commensurate structure every second potential well is occupied by an atom and that the misfit $m_\textrm{AA} = b(\textrm{A,A})-2= -0.1$
is negative. This value of misfit is larger than in the case of Kr/Graphite system, for which it is equal
to about $-0.06$ \cite{Ar-Kr-my}, but assuming the value of the misfit equal to $-0.06$, the FK model
leads to the stable commensurate phase even for unreasonably low amplitudes of the surface field.
To model Ar atoms we assumed that $v_\textrm{B}=0.003$, $b(\textrm{B,B})=1.8$ and $K(\textrm{B,B}) = 0.9$
The parameters $K(\textrm{A,B})$ and $b(\textrm{A,B})$ are assumed to be given by the Lorentz-Berthelot mixing rules, so that
\begin{equation}
K(\textrm{A,B})=\sqrt{K(\textrm{A,A})K(\textrm{B,B})}  \qquad \text{and} \qquad b(\textrm{A,B})=0.5\left[b(\textrm{A,A})+b(\textrm{B,B})\right].
\label{eq:Konmix}
\end{equation}

We  considered the chains that begin and terminate with
 atoms B. This assumption arises directly from the observation that in two-dimensional systems,
the domain-wall
structures appearing only after
the peripheries of the cluster made of Kr atoms  are covered with a filled single layer of Ar atoms.
The situation in which argon atoms are located only at the ends of the chain corresponds to
the model labeled as Model I. We considered two versions of this model assuming that there is only
one argon atom at each chain end (Model I1) and that each end is decorated by two argon atoms (Model I2).
Then, in the Model II, we assumed that at a certain position after the atom $N_1$, i.e., at the position $N_1+1$,
a mixed wall of the thickness
equal to $k$ atoms appears. The version of the model with a given value of $k$ is named Model II$(k,N_1)$.

In order
to consider mixed walls, we  assumed that within the wall, the parameters representing the elastic constant, $K_w$,
the equilibrium distance, $b_w$, and the amplitude of the substrate potential, $v_w$, are all dependent on the
wall composition measured by the mole fraction of component B, $x_\textrm{B}$, and  are given by:
\begin{equation}
K_w = x_\textrm{B}K(\textrm{B,B})+(1-x_\textrm{B})K(\textrm{A,A}),
\label{eq:Kon2}
\end{equation}
\begin{equation}
b_w=x_\textrm{B}b(\textrm{B,B})=(1-x_\textrm{B})b(\textrm{A,A})
\label{eq:Kon3}
\end{equation}
and
\begin{equation}
v_w=x_\textrm{B}v_\textrm{B}+(1-x_\textrm{B})v_\textrm{A}\,.
\label{eq:Kon4}
\end{equation}

The pairs $(N_1,N_1+1)$ and $(N_1+k,N_1+k+1)$, i.e., at both sides of the wall, consist of one A atom and one wall atom,
so that the respective elastic constants and equilibrium distances are given by
\begin{equation}
K(\textrm{A},w) =x_\textrm{B}K(\textrm{A,B})+(1-x_\textrm{B})K(\textrm{A,A}) \qquad \text{and} \qquad b(\textrm{A},w) = x_\textrm{B}b(\textrm{A,B})=(1-x_\textrm{B})b(\textrm{A,A}).
\label{eq:Konmix2}
\end{equation}
The above assumption takes into account the fact that when the wall is made only of atoms B, the
interaction between
the pairs $(N_1,N_1+1)$ and $(N_1+k,N_1+k+1)$ is the same as between the pairs (A,B) and $(N-1,N)$, which are
equal to $K(\textrm{A,B})$ and $b(\textrm{A,B})$, respectively.

Of course, the above assumed additivity relations are too simple to properly describe the effects of wall
composition  on the
 behavior of adsorbed films, but we will demonstrate that even such a simple model leads to the results
qualitatively very similar to those obtained for two-dimensional films.

Then, we introduced the displacements $u_i = x_i/b-2i$, with $b$ being the equilibrium distance for the pair $i$
and $i+1$, and rewrote equation (\ref{eq:Kon1}) in the form
\begin{equation}
E = \frac{1}{2}\left\{\sum_{i=1}^{N-1}\hat{K}_{i,i+1}\left[u_{i+1}-u_{i}-m_{i,i+1}\right]^2 + \sum_{i=1}^N\hat{v}_{i}\left[1-\cos(2\pi u_i)\right]\right\},
\label{eq:Kon5}
\end{equation}
where the energy is expressed in units of $K(\textrm{A,A})a^2$, and $\hat{K}_{i,i+1}=K_{i,i+1}/[K(\textrm{A,A})a^2]$ and
$\hat{v}_i=v_i/[K(\textrm{A,A})a^2]$ and the misfits $m_{i,i+1}$ are defined as $m_{i,i+1}=b(\alpha,\beta)/a-2$ with
$(\alpha,\beta) = (\textrm{A,A})$, $(\textrm{A,B})$, $(\textrm{A},w)$ or $(w,w)$.

In the case of Model I, the energy given by equation (\ref{eq:Kon5}) depends on the number of argon atoms at the
ends of the chain, while in the Models II$(k,N_1)$ it depends on both the thickness of the wall ($k$) and
its position ($N_1$).

\begin{figure}[!b]
\centerline{
\includegraphics[scale=0.39]{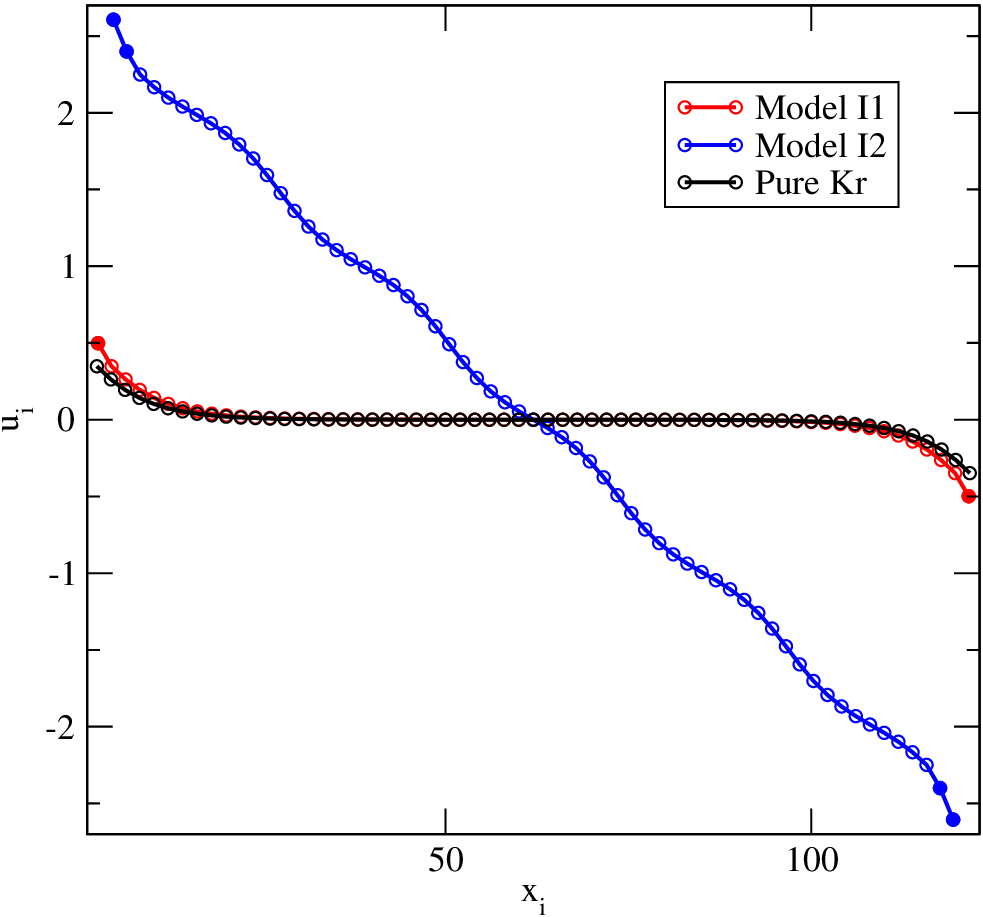}
}
\caption{(Color online) Atomic displacements versus atomic positions for finite chains consisting of 41 atoms. Black circles and line
correspond to pure chain of B atoms, while red and blue circles and lines correspond to the chains decorated by one and two
atoms A at each end of the chain, respectively.}
\label{fig7}
\end{figure}

One should note that the systems considered in the previous section were studied using Monte Carlo method allowing for the exchange of atoms identity,
so that the distances between the patch boundary and the walls and the wall-wall distances inside the
patch correspond to the equilibrium states. In the present one-dimensional
model, the location of the wall is predetermined by $N_1$, and one needs to minimize the energy with respect to all
displacements $\{u_i\}$ as well as with respect to $N_1$ in order to find the equilibrium state of the chain.
On the other hand, the wall thickness and composition are controlled by the argon mole fraction in the
chain.
The first series of calculations is aimed at determining whether the assumed model predicts the formation
of solitons when B atoms are located only at the chain boundaries, i.e., when $x_\textrm{B}=0.0$ within the wall.
The calculations carried out for different chain lengths, with $N$ ranging from 11 to 101, and using Model I1 always led to a single commensurate domain, i.e., to the same qualitative results as for pure A chain.
The only effect
of the B atoms located at the chain ends was an increase of atomic displacements near the chain ends
(see figure~\ref{fig7}). This result agrees very well with two-dimensional Monte Carlo simulations carried out
for low $x_\textrm{Ar}$, up to the value for which the entire boundary of commensurate Kr patch was covered by a single
layer of Ar atoms.
However, when the Model I2 was used,
the introduced perturbation appeared to be sufficiently high to lead to the formation of solitons, i.e., the
domain-wall structure (see figure~\ref{fig7}).
From the results of the previous section it follows that already at very low temperatures, the Monte Carlo simulation,
starting with the configuration containing two rows of argon atoms at the boundaries, leads to the formation
of a wall consisting of two rows of Ar atoms inside the Kr patch, and the boundaries are decorated by only a single layer
of Ar atoms. Therefore, it is quite likely that the structure with two B atoms  at each end of the chain is not
the equilibrium state for the chain with four B atoms.

Then, we considered the systems described by Models II$(k,N_1)$ with the wall of a thickness $k$
located inside the chain after the atom $N_1$. At first, we  assumed that the chain length is fixed
($N=41$) and studied the behavior of the model when the concentration of B atoms within the wall changes.

This situation mimics finite submonolayer systems with the gradually increasing concentration of argon.
In the two-dimensional systems, the number of atoms is proportional to $N^2$. Our
choice of $N$ corresponds to clusters with the number of atoms equal to about $1500\pm 100$ i.e., not much
larger than those with 1333 atom that were studied by Monte Carlo simulations.

\begin{figure}[!t]
\centerline{
\includegraphics[scale=0.39]{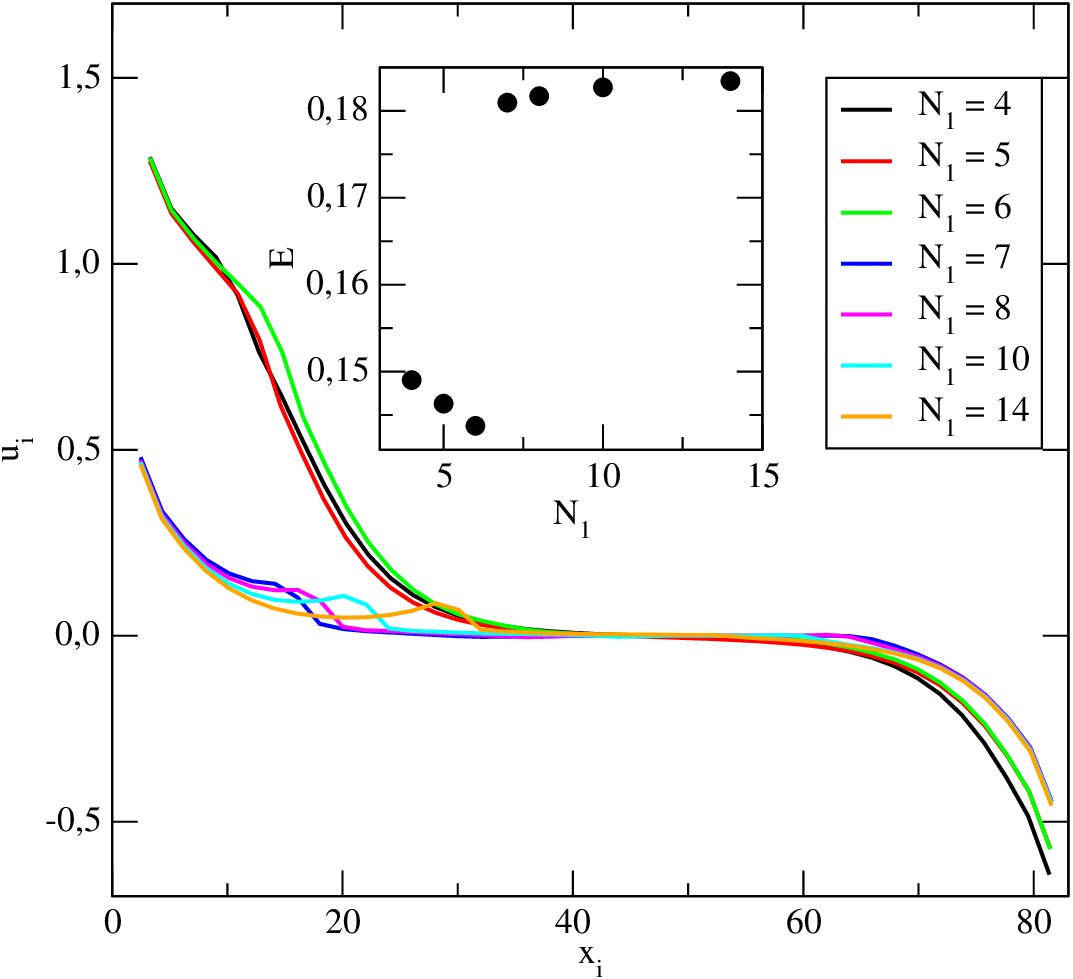}
}
\caption{(Color online) The main figure shows atomic displacements versus atomic positions for finite chains consisting of 41 atoms
with the ends decorated with one A atom and containing one A atom within the chain at the position $N_1+1$ (given in the legend
and marked by a filled circle). The inset shows the changes of the system energy with the position of atom A within the chain.}
\label{fig8}
\end{figure}

The calculations carried out within the Model II($1,N_1$) for the wall consisting entirely of argon
($x_\textrm{B} =1.0$) demonstrated that when $N_1\leqslant 6$, there develops a single soliton located behind the wall (see figure~\ref{fig8}). The energy of the system acquires a minimum value when $N_1=6$ (see the inset a to figure~\ref{fig8}).

When $N_1>6$, a single
atom B inside the chain does not lead to the formation of a wall. Only a rather small perturbation appears
close to the impurity atom
 (figure~\ref{fig8}). Evidently, the commensurate stripe between the boundary and the
atom B is already wide enough to sustain the stress induced by the impurity.

\begin{figure}[!t]
\centerline{
\includegraphics[scale=0.39]{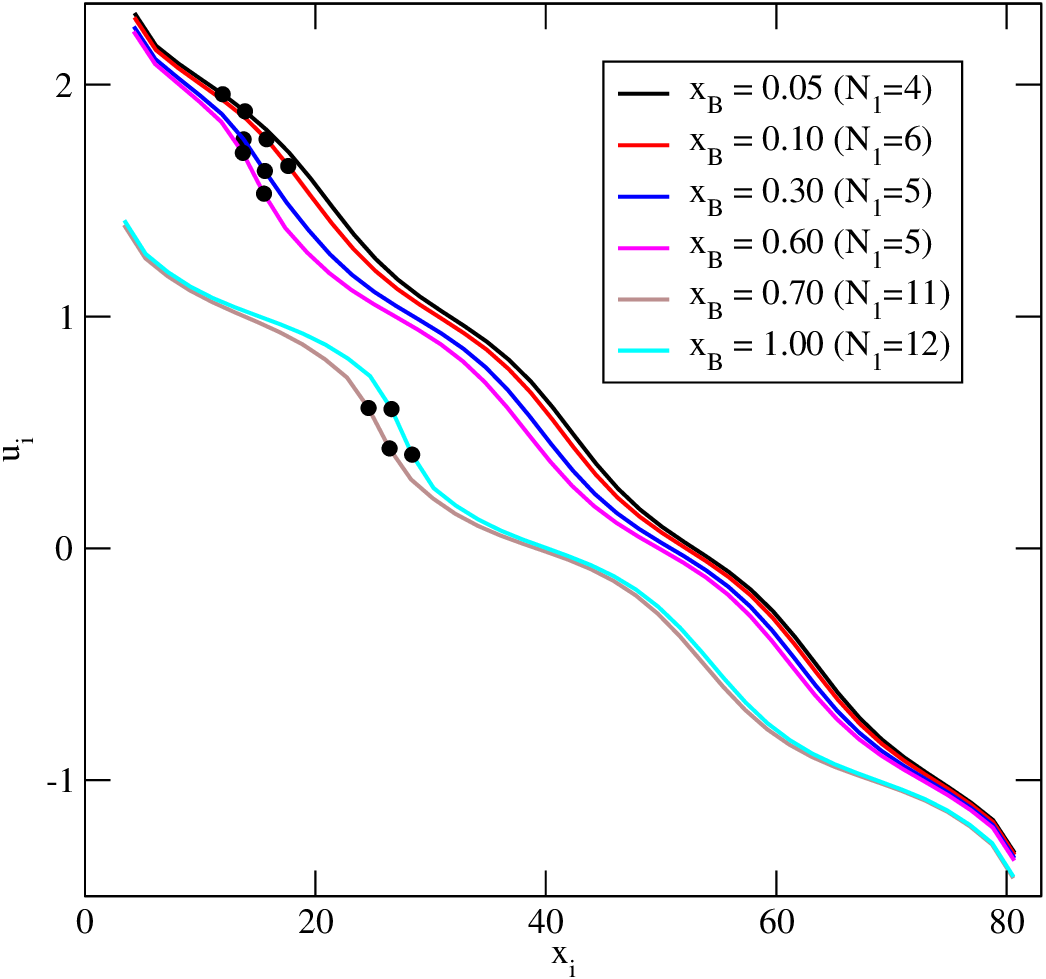}
}
\caption{(Color online) Atomic displacements versus atomic positions for finite chains consisting of 41 atoms with the ends
decorated with one A atom and containing two A atoms within the chain at the positions marked by black circles obtained
for different compositions of the wall inside the chain and corresponding to the location of the wall that minimizes the
system energy.}
\label{fig9}
\end{figure}

In the case of the Model II($2,N_1$) with $N=41$, we found out that for $x_\textrm{B}$ up to 0.04, the
commensurate structure is a stable state for all possible values of $N_1$. However, already
for $x_\textrm{B}=0.05$, the system undergoes
a transition to the domain-wall structure. Figure~\ref{fig9} shows the atomic displacements obtained for
a series of systems with $x_\textrm{B}$ between 0.05 and 1.0.
Thus, the results indicate that even when the perturbation introduced by a small concentration of component B
within the wall is weak, it leads to the development of solitons. Two types of behavior can be singled out.
For $x_\textrm{B}\in(0.05, 0.6)$, there are three solitons and with the increase of $x_\textrm{B}$, the B atoms gradually
approach the position of the first soliton.
In the second regime, only two solitons appear and the B atoms are located right within the wall.

Of course, the results obtained for a high concentration of atoms B in the wall, do not correspond well
to the 2D systems studied using Monte Carlo simulations. For a large concentration of atoms B, one would
expect the formation of several mixed walls with lower values of $x_\textrm{B}$ rather than the appearance of a single
wall with a high value of $x_\textrm{B}$.  Nevertheless, the model presented catches the most important features
of the C-IC transition in finite systems driven by the presence of impurities.

\section{Summary}

Using Monte Carlo simulation method we  studied the C-IC transition
which occurs at low temperatures in
finite clusters of Kr atoms contaminated with argon.
It is demonstrated that
pure krypton clusters form only the commensurate structure. This is consistent with experimental data clearly
demonstrating that the commensurate structure is stable as long as the density does not exceed the density
of a fully covered commensurate phase \cite{krgr4}. Also, the clusters retain a commensurate structure when small amounts
of argon are added to the system. It is established that argon atoms are predominantly located along the
cluster boundaries and the C structure is stable up to the point when the entire cluster boundary is covered with
a single layer of argon atoms. As soon as the argon concentration becomes higher than this threshold value, the
system undergoes the C-IC transition leading to the formation of the domain-wall structure. The walls separating
the commensurate domains are found to be heavy and their composition to change with the argon concentration.
In particular, we found out that the argon concentration within the wall changes linearly with the total argon mole
fraction in the system. Thus, at a certain upper limit of the argon concentration in the film, the walls are made
of argon only, while the cluster boundaries are still covered with a single atomic layer of argon. A further increase
of argon concentration causes the walls made of argon and the argon layer along the patch boundaries to become
thicker. It should be noted that the size of commensurate domains does not depend much upon the argon concentration
and the domains are made of pure krypton. This demonstrates that the mixing does not occur in the C phase. On the other
hand, the mixing does occur within the incommensurate walls over a rather wide range of argon concentration.

The above described behavior can be attributed to the rather large changes in the interaction energies between different
pairs (Kr-Kr, Ar-Kr and Ar-Ar) when the system undergoes the C-IC transition. In the C phase, the distance between
neighboring atoms is determined by the structure of the graphite lattice. A considerably smaller misfit of krypton compared to argon, causes the system to maximize the number of Kr-Kr pairs and to avoid the appearance of Ar-Kr and
Ar-Ar pairs. In the incommensurate state, i.e., within the walls, the energy of interaction between Ar-Kr and Ar-Ar
pairs is considerably higher since the distances between nearest neighbors within the heavy walls are considerably
lower and closer to the locations of the LJ potential minima. This effect enhances the tendency toward mixing. Besides,
the stability of walls is further enhanced by rather large entropic effects.

We  also showed that similar C-IC transition emerges from a simple one-dimensional Frenkel-Kontorowa model applied
to finite mixed chains. It is demonstrated that for appropriately chosen parameters of the model, its
behavior is qualitatively very similar to that obtained from Monte Carlo simulations of two-dimensional systems.

\ukrainianpart

\title{Механізм формування структури доменної стінки в субмоношарових плівках Ar-Kr на графіті}

\author{A. Патрикієв, С. Соколовський}

\address{Відділ моделювання фізико-хімічних процесів, хімічний факультет, Університет імені Марії Склодовської-Кюрі, 20031 Люблін, Польща}

\makeukrtitle

\begin{abstract}
\tolerance=3000%
Використовуючи метод моделювання Монте-Карло в канонічному ансамблі, ми дослідили
перехід від співвимірної до неспіввимірної фази у двовимірних скінчених змішаних кластерах  Ar і Kr,
адсорбованих на графітній базисній площині при низьких температурах. Показано, що перехід відбувається, коли концентрація аргону
перевищує значення, необхідні для покриття периферій кластера. Неспіввимірна фаза демонструє
подібну структуру доменної стінки до тої, яка була спостережена у плівках чистого криптону при густинах, що перевищують
ідеальну $(\sqrt{3}\times\sqrt{3})R30^\circ$  співвимірну фазу,
але розмір співвимірних доменів  значно не змінюється зі зміною розміру кластера.
Коли концентрація аргону зростає, склад доменних стінок змінюється, в той час як співвимірні домени складаються з чистого криптону.
Ми побудували просту одновимірну модель Френкеля-Конторової, яка приводить до результатів, що добре якісно узгоджуються
з результатами Монте-Карло, отриманими для двовимірних систем.
\end{abstract}

\end{document}